\documentclass[epj,referee]{svjour}
\usepackage{graphicx}
\begin{document}
\title{The structural role of weak and strong links in a financial market network}
\author{Antonios Garas\inst{1}, Panos Argyrakis\inst{1}, \and Shlomo Havlin\inst{2}
}                     
\offprints{Panos Argyrakis}          
\institute{Department of Physics, University of Thessaloniki, 54124 Thessaloniki, Greece \and Minerva Center and Department of Physics, Bar-Ilan University, Ramat Gan 52900, Israel}
\abstract{
We investigate the properties of correlation based networks originating from economic complex systems, such as the network of stocks traded at the New York Stock Exchange (NYSE). The weaker links (low correlation) of the system are found to contribute to the overall connectivity of the network significantly more than the strong links (high correlation). We find that nodes connected through strong links form well defined communities. These communities are clustered together in more complex ways compared to the widely used classification according to the economic activity. We find that some companies, such as General Electric (GE), Coca Cola (KO), and others, can be involved in different communities. The communities are found to be quite stable over time. Similar results were obtained by investigating markets completely different in size and properties, such as the Athens Stock Exchange (ASE). The present method may be also useful for other networks generated through correlations.
} 
\authorrunning{Antonios Garas et al.}
\maketitle
\section{Introduction}
\label{intro}
Recently there has been a growing interest to better understand complex systems. A complex system is generally composed of many interacting elements in various ways. A network representation is found useful to characterise the system, by associating each element by a node and each interaction by a link (weighted or not). To understand the network structure and function, various tools from statistical physics have been developed. These tools, such as scaling theory, percolation, and fractal analysis \cite{book:PastorSatorras,book:Mendes,bb:WattsStrogatz,bb:AlbertBarabasi,bb:SongHavlinMakse,bb:Serrano,bb:Clauset,bb:Eriksen,bb:Cohen,bb:Braunstein}, enable us to extract useful information and to better describe properties of complex systems. Examples of complex systems that have been recently investigated from this perspective include the Internet \cite{bb:Faloutsos,bb:ShaiCarmi}, the World Wide Web \cite{bb:AlbertJeongBarabasi}, communication networks \cite{bb:PastorVespignani}, food webs \cite{bb:Garlaschelli}, sexual contact networks \cite{bb:Liljeros} and economic networks \cite{bb:Mantegna,bb:Tumminello,bb:OnnellaKaskiKertesz}. 

The problem of extracting useful information from a system becomes more difficult in the case of correlation based networks, since these networks are usually complete graphs (all links between elements are present).
On the other hand, understanding the behavior of networks originating from empirical correlation matrices is a very important task in many scientific fields, since correlation matrices appear in the study of multivariate time series.
In order to make correlation based networks simpler to understand and extract information from them, the use of filtering techniques was suggested~\cite{bb:Mantegna,bb:Tumminello}. Filtering techniques, like the Minimum Spanning Tree (MST) \cite{bb:Mantegna}, reduce the number of links of the network and keep all the nodes connected with a total maximum weight.

In this work we study correlation based networks by exploring the evolution and temporal dynamics of the structures occurring after the removal of a certain fraction $q$ of links, without forcing all the nodes to remain connected to the network. We identify a particular value of this fraction, $q \approx 0.995$, close to which structural properties of the network become clearer. We therefore study the communities of stocks at this particular point.

\begin{figure*}
\centering
\includegraphics[width=11cm]{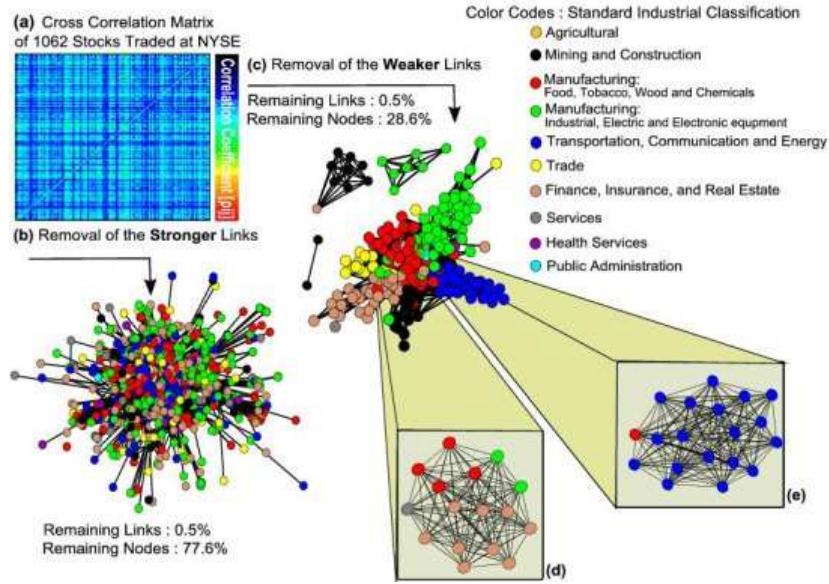}

\caption{Visualization of the different network structures that occur after removing 99.5\% of the links of the initial fully connected network. (a)~Pictorial representation of a cross correlation matrix of a portfolio of 1062 stocks traded at the NYSE. 
(b)~The network after removing the stronger links of the initial fully connected network. (c)~The network after removing the weaker links of the initial fully connected  network. (d)~A community of 16 stocks, belonging to 3 overlapping cliques of 14 elements. The tick names of the stocks forming this community are: AIG(tan), AXP(tan), BMY(red), C(tan), CL(red), DIS(gray), EMR(green), FNM(tan), GE(green), JPM(tan), KO(red), MER(tan), MMC(tan), SGP(red), TA(tan) and TY(tan). (e)~A community of 19 stocks, belonging to 3 overlapping cliques of 17 elements. The tick names of the stocks forming this community are: AEP(blue), AIT(blue), BEL(blue), BGE(blue), BLS(blue), CPL(blue), CSR(blue), D(blue), DUK(blue), ED(blue), FPL(blue), GTE(blue), KO(red), NSP(blue), PCG(blue), PEG(blue), SBC(blue), SO(blue) and USW(blue). The color codes that we use are according to the Standard Industrial Classification (SIC) system for classifying industries. }
\label{fig:1}
\end{figure*}

\section{Methods:~Creating and Destructing the Network}
\label{sec:1}

In many complex systems the network is built by using correlations between the dynamics of the nodes. This is, for example, the case in economic networks, where a weighted link is assigned between two nodes representing different stocks according to the cross correlation between the return time series of each stock.

In the present study we create a correlation based network using the closing prices of a portfolio comprising of 1062 stocks traded in the New York Stock Exchange (NYSE) in the period 1987 to 1998. From the daily closing price time series we can create a correlation based network by following the procedure that we describe bellow.
First we calculate the correlation coefficient between each pair of stocks, $i$ and $j$ defined as,
\begin{equation} \rho_{ij}=\frac{\left<r_{i}r_{j}\right>-\left<r_{i}\right>\left<r_{j}\right>}{\sqrt{\Big{(}\left<r_{i}^2\right>-\left<r_{i}\right>^2\Big{)}\Big{(}\left<r_{j}^2\right>-\left<r_{j}\right>^2\Big{)}}}
\label{eq:rho}
\end{equation}
where $\left<\dots\right>$  is the time average over the investigated time period. Here $r_{i}=r_{i}(t)$ is the logarithmic return, defined by $r_{i}(t)=\mbox{ln}P_{i}(t)-\mbox{ln}P_{i}(t-\Delta t)$, and $P_{i}(t)$ is the daily closing price of stock $i$ at day $t$. If two stocks, $i$ and $j$, are completely correlated (anti-correlated) then $\rho_{ij}=+1(-1)$, while if the two stocks are completely uncorrelated then $\rho_{ij}=0$\footnote{ This is true for linearly correlated time series.}. In our case $\Delta t = 1$ day. By calculating the correlation coefficient for all pair of stocks, we obtain the correlation coefficient matrix of the system. Such matrices were studied in~\cite{bb:Laloux-RMT,bb:Plerou-RMT} and are known to have a large amount of noise, that can be attributed to false correlation estimates due to the finite size length of the time series. 

An empirical correlation matrix can be viewed as a fully connected weighted network by transforming the correlation coefficient to a distance, using an appropriate function as a metric~\cite{bb:Mantegna}. The function that we used for this transformation is~\cite{bb:Mantegna}
\begin{equation} 
d_{ij}=\sqrt{2\left(1-\rho_{ij}\right)},\ \ \ \ 0\leq d_{ij}\leq 2,
\label{eq:d}
\end{equation}
where small values of the distance $d_{ij}$ imply strong correlation for the pair of stocks, $i$ and $j$, and vice versa.

Next we investigate two methods of removing the links from this fully connected network, both resulting in sparser graphs with totally different properties. From these differences we can learn about the structure of the network. We begin by sorting the weights in increasing order. In the first method we repeatedly remove links from lower to higher values of $d_{ij}$ (high to low correlations). In the second method we repeatedly remove links starting from the highest values to lowest values of $d_{ij}$ (low to high correlations). A similar approach to our second method was implemented by Onnela et al.  \cite{bb:OnnellaKaskiKertesz}. In this work the authors used two sets, one of 116 and one of 477 stocks traded in the NYSE, and they begun adding links to the initially completely disconnected network starting from the highest to lowest correlation values.

We find that after removing the stronger correlated links of the network (links with low $d_{ij}$ values), the network remains connected until we have removed almost 99\% of its original connections. On the contrary when we remove links starting from the weakest correlated (links with high $d_{ij}$ values) the network starts to lose its nodes much earlier, after removing only about 30\% of its original weakest links (see Fig.~\ref{fig:4}a). This result suggests that strong links and weak links play very different roles in the topology. While strong links are usualy situated in possition which increases local connectivity within the communities, the weak links contribute more to the global connectivity, i.e. in connections between communities. An interesting observation is that in both cases the disintegration of the network takes place gradually, mainly because some very small clusters become disconnected from the largest cluster.

In Figure~\ref{fig:1} we plot some representative results of the above procedure. Fig.~\ref{fig:1}(a) shows a representation of a cross correlation matrix of a portfolio of 1062 stocks traded at the NYSE. In Fig.~\ref{fig:1}(b) we draw the network left after removing 99.5\% of the stronger links of the initial fully connected network and in Fig.~\ref{fig:1}(c) we draw the network left after removing the same fraction of the weaker links. From this figure it is apparent that the two networks, even by visual investigation, are totally different. When we remove the stronger links we get a network that has more nodes but less structure, while when we remove the weaker links the resulting network has fewer nodes, but these nodes are clustered together mainly in accordance to the sector of economic activity. In order to classify the stocks into different sectors we used the Standard Industrial Clasification (SIC) system for the classification of industries~\cite{bb:SIC}.

\begin{figure}
\resizebox{1\columnwidth}{!}{%
\includegraphics{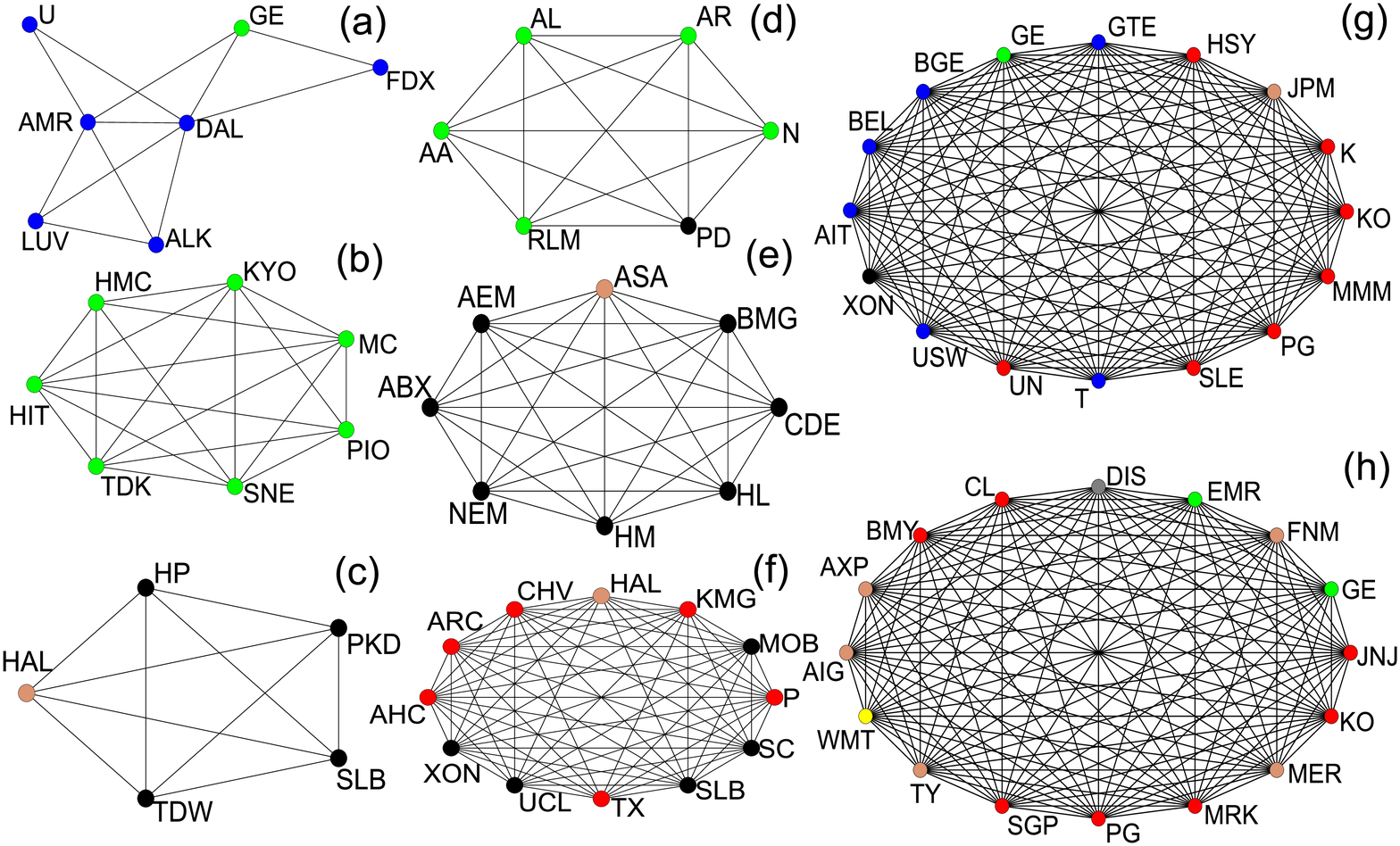}
}
\caption{Further examples of communities of stocks that we identified using the $k$-clique method to the network that remains after removing 99.5\% of the links of the weaker original fully connected network.}
\label{fig:2}
\end{figure}

To further investigate the way these clusters are interconnected we apply the $k$-clique method to detect communities~\cite{bb:Vicsek} after removing q=0.995 of the weak links. The reason for choosing this value will be clear later when we find that close to this value structural properties of the communities are clearer to see. A maximal complete subgraph of a network is called clique. In addition, a smaller complete subgraph with $k$ nodes,  that in general can be included in a larger one, is called $k$-clique. In a network, a large complete subgraph of size $s$, $(k\leq s)$ contains $\left( \begin{array}{c} s \\ k \end{array} \right) $ different smaller complete subgraphs of size $k$ ($k$-cliques). The algorithm we are using~\cite{bb:Vicsek} is able to calculate all the $k$-cliques of the network, if the network is sparse enough, and therefore it allows us to identify a wealth of communities of stocks. Examples of such communities are shown in Figure~\ref{fig:1}(d),~Figure~\ref{fig:1}(e) and Figure~\ref{fig:2}. 

\begin{figure}
\resizebox{1\columnwidth}{!}{%
\includegraphics{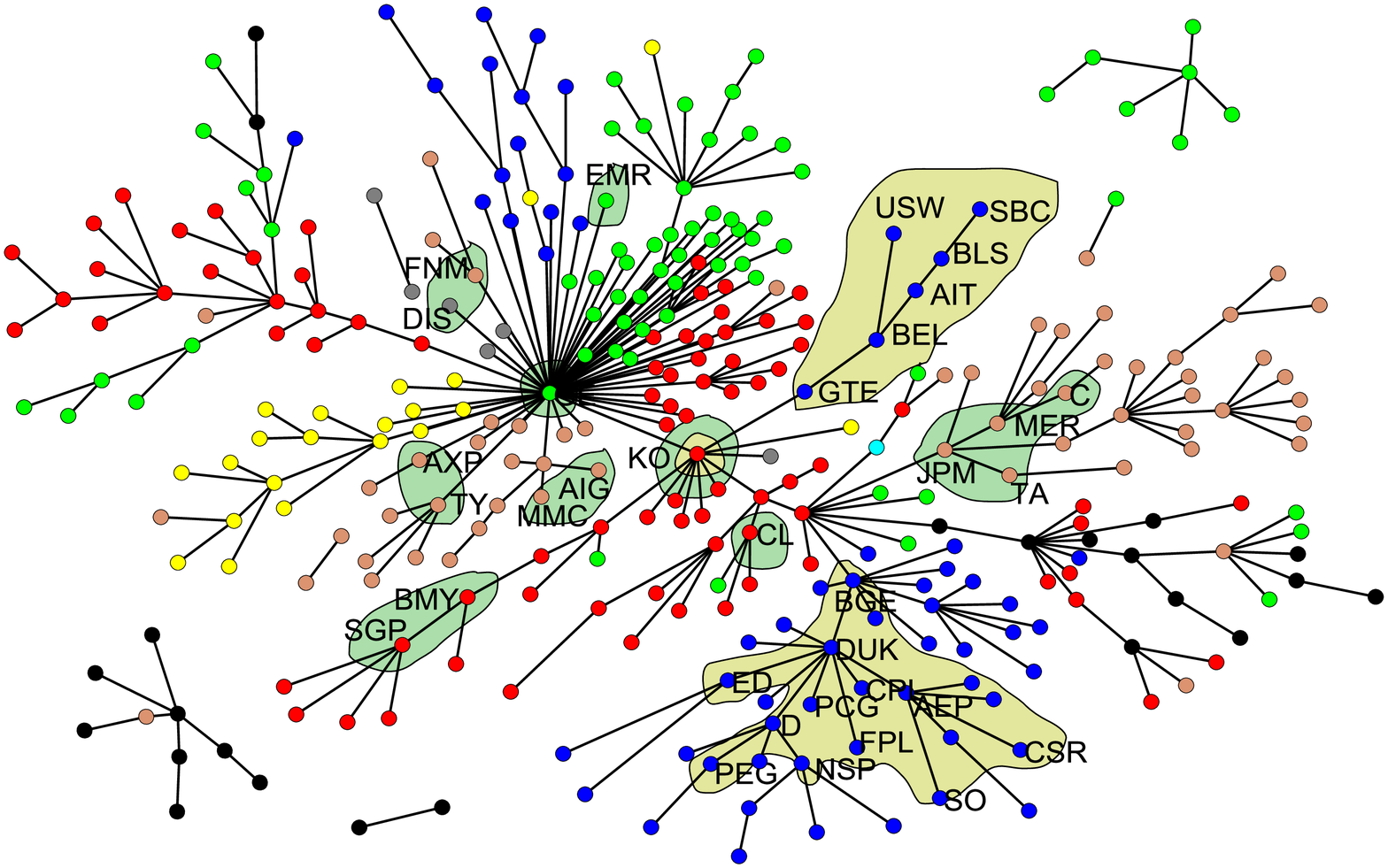}
}
\caption{Minimum Spanning Tree (MST) of the connected nodes of the network after removing 99.5\% of the weaker links of the initially fully connected  network. The nodes that are marked here with their corresponding symbols are the nodes belonging to the two examples of communities that were described in Fig.~\ref{fig:1}.}
\label{fig:3}
\end{figure}

\section{Results}
\label{sec:2}

The first finding is that when we remove the weaker links, the nodes are clustered according to the sector of economic activity, and this is in general expected, since the intra-sector correlations between stocks are usually very strong. Typical mean values of weights of links classified as intra-sector, inter-sector, intra-sub-sector, etc. and their estimated standard deviation were computed by Tumminello et al. \cite{bb:TumminelloChaos} using a bootstrap technique. However, with our approach  we can identify also, in the same community, nodes from different sectors. As an example, in Figure~\ref{fig:1}(d) we find a very well-connected community of the stocks  AIG(tan), AXP(tan), BMY(red), C(tan), CL(red), DIS(gray), EMR(green), FNM(tan), GE (green), JPM(tan), KO(red), MER(tan), MMC(tan), SGP (red), TA(tan) and TY(tan). As can be seen from the color code, these stocks belong to 4 different sectors. The meaning of this result is probably related to other activities of the companies that are not reflected by their main sector classification, but they affect the performance of the stocks in a non trivial way.

In Fig~\ref{fig:2} we present a variety of further communities of stocks belonging to one or more sectors of economic activity. Most of these communities are almost fully connected subgraphs of the network. From Table~\ref{tab:communities}, where the tick names of each community's stocks  are listed, we can identify several stocks that belong to more than one community. This finding of overlapping communities is novel and very important since it points that there is a number of stocks that can influence many other stocks or even group of stocks, belonging to different sectors, and vice versa. Examples of such stocks that we identify from Table~\ref{tab:communities} are: General Electric (GE), Coca Cola (KO), Exxon Mobil (XON) and Procter \& Gamble (PG). All the above examples refer to very large capitalization companies that point towards the significance of such blue chips to the overall market activity.

Another widely used method to identify clusters in a network of $n$ of stocks according to the sector of economic activity is the Minimum Spanning Tree (MST) technique \cite{bb:Mantegna,bb:OnnellaKaskiKertesz,bb:Onnela,bb:Garas}.  An MST analysis was performed using a dataset similar to the one we use by Bonanno et al. \cite{bb:Bonanno}. This technique filters the original correlation matrix and keeps only a tree of $n-1$ links out of the original $n(n-1)/2$ links with total minimal distance. Application of this method to the original fully connected network or to the network given in Fig~\ref{fig:1}(c), yields a nice clustering of stocks according to their economic activity, but the communities we identified in Fig~\ref{fig:1} are fragmented and, as shown in Fig~\ref{fig:3}, this fragmentation is more pronounced for stocks belonging to different economic sectors.

\begin{figure*}
\centering
\resizebox{1.5\columnwidth}{!}{%
\includegraphics{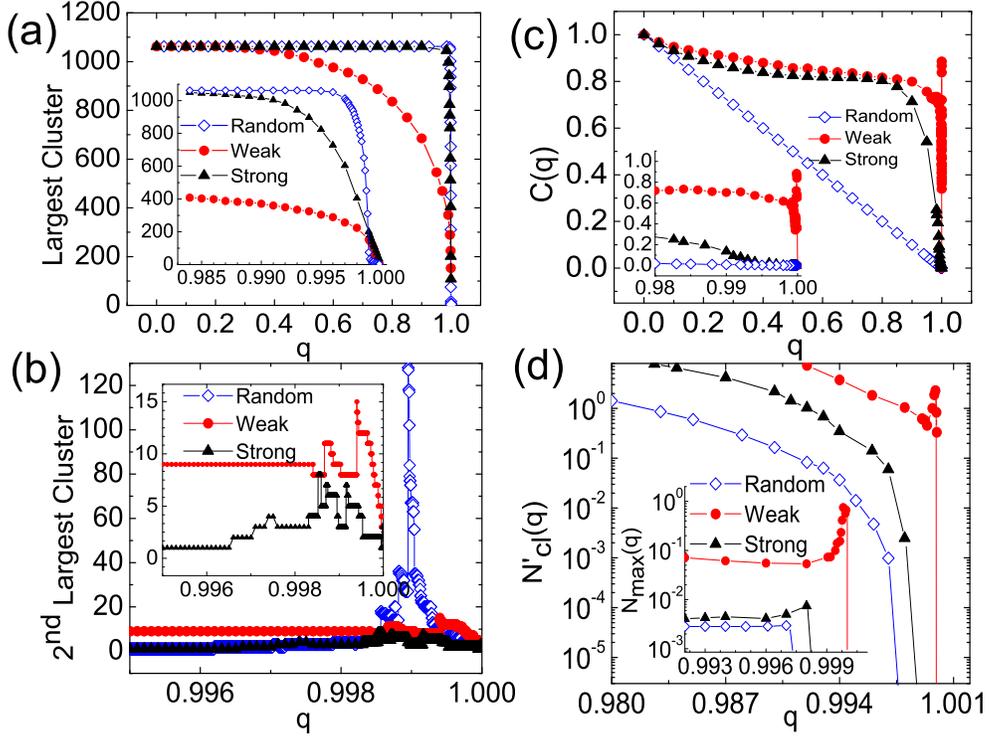}
}
\caption{(a)~The number of nodes belonging to the largest cluster of the network versus the fraction of removed links $q$. Inset:~A zoom of the area around $q=0.99$. (b)~The number of nodes belonging to the second largest cluster of the network versus the fraction of removed links $q$. Inset:~A zoom of the area around $q=0.99$. 
(c)~Clustering coefficient $C(q)$ for the three different cases of the link removal procedure. Inset:~A zoom of the area around $q=0.99$.
(d)~The relative number of cliques $N'_{\mbox{cl}}(q)$, after we removed a fraction $q$ of links versus the fraction of removed links $q$. Inset:~The relative number of maximum  clique size $N_{\mbox{max}}(q)$ after removing a fraction $q$ of links versus the fraction of removed links $q$.}
\label{fig:4}
\end{figure*}

Next we focus on other properties of the network. In Fig~\ref{fig:4}(a) we compare the largest cluster of the network versus the fraction of removed links $q$ using weak removal, strong removal and random removal. For random removal it is seen that the network remains connected until we have removed over 99.9\% of the original links. Indeed, after we remove 99.9\% of the original links the value of the parameter $\kappa$ \cite{bb:Cohen} becomes $\kappa=\left< k^2 \right>/\left< k \right>=2, $ where $k$ is the degree of the nodes and $\left< \dots \right>$ is the average over all the nodes. At this point there is a percolation transition and the network breaks into clusters of connected components. As we can see from Fig.~\ref{fig:4}(b) the second largest cluster has a maximum size for $q=0.999$ and its size is comparable to the size of the original network. This behavior is completely different from the way the network disintegrates when we remove first its stronger or its weaker links. For both these cases, as we remove the links of the network some isolated nodes, or even some small clusters of nodes, gradually losing all their links and they are being removed from the network, therefore, the network is being stripped of its nodes and it becomes disconnected without a sharp percolation transition. The desintegration of the network is faster when we remove the weaker correlated links. This result shows that the weak links are responsible for the global connectivity of the network, while most of the strong links form local structures.

For the analysis that follows we calculated further properties of the remaining connected component of the network as a function of the fraction of removed links $q$ by applying all three different removal procedures we described.

A property that plays important role in the structure and connectivity of many different kind of networks is the formation of cliques. This property can be easily understood for the case of social networks where it represents circles of friends or acquaintances in which every member knows well every other member of the clique, but usually does not know members of other cliques. One method to quantify the tendency to cluster in this way is the clustering coefficient $C(q)$ \cite{bb:WattsStrogatz,bb:AlbertBarabasi}, that is defined as follows. If a vertex $i$ has $k_{i}$ neighbors then at most $k_{i}(k_{i}-1)/2$ edges can exist between them (this occurs when every neighbor of $i$ is connected to every other neighbour of $i$). Let $C_{i}(q)$ denote the fraction of such existing edges for node $i$, then $C(q)$ is defined as the average of $C_{i}(q)$ over all connected nodes of the network.

We used the clustering coefficient $C(q)$ to compare the connectivity of the network structures that survive after removing a fraction $q$ of the original links, for the three different cases of the link removal procedure mentioned above. The results of this analysis are shown in Figure~\ref{fig:4}(c). Note that the network for which its strongest links survive, it has always higher clustering coefficient, again showing that strong links make more connections locally, while weak links are responsible for the global network structure.

In addition to the clustering coefficient, there exist another useful quantity that can yield more information on the structure of the network. This is the total number of cliques and their size. We calculate the \textit{relative number of cliques} $N'_{\mbox{cl}}(q)$, defined as the total number of cliques that exist in the network divided with the number of its nodes, \[ N'_{\mbox{cl}}(q)=N_{\mbox{cl}}(q)/N_{\mbox{nodes}}(q). \] We also calculate the \textit{relative number of maximum  clique size} $N_{\mbox{max}}(q)$, defined as the ratio between the maximum clique size $\mbox{Max}_{\mbox{cl}}(q)$ in the network and the number of nodes, \[ N_{\mbox{max}}(q)=\mbox{Max}_{\mbox{cl}}(q)/N_{\mbox{nodes}}(q). \] Finding whether there is a clique of a given size in a graph is a NP-complete problem. We thus studied the behavior of the above quantities only for the range close to $q=0.99$, where the network is very sparse, but it is enough to help us to draw some interesting conclusions. Results of this analysis are shown in Figure~\ref{fig:4}(d).

From Figures~\ref{fig:4}(c) and~\ref{fig:4}(d) we can see that the remaining network after removing the stronger links although it has more nodes compared to the network obtained after removing the same amount of the weaker links, has a much lower internal structure (it has fewer and smaller cliques). This means that inside the original network there exist a well defined underlying structure of strongly connected components, and a bulk of weaker, less meaningful connections. The more weaker links we remove from the network, the more visible this structure becomes. This is the reason behind the sharp peaks in Figures~\ref{fig:4}(c) and ~\ref{fig:4}(d). This increase of $C(q)$, $N'_{\mbox{cl}}(q)$ and $N_{\mbox{max}}(q)$ with $q$ that starts to occur around $q=0.998$ only when removing weak links suggests that at this regime we are able to uncover some of the most important structural features of the network. This justify our analysis of the communities in Figs~\ref{fig:1} and~\ref{fig:2} at values of $q=0.995$. We find similar results for $0.995<q<0.999$.

Next, we study the dynamic evolution of the networks by comparing links using annual data. We approach this by analysing the network in a similar fashion to the analysis of the dynamics of Minimum Spanning Trees~\cite{bb:Onnela,bb:Garas}. The \textit{single-step similarity probability} is a measure of how many common links exist in the networks for two consecutive years, after we removed the same percentage of removed links $q$. The single step similarity probability is defined as:
\begin{equation}
	S^{q}(t)=\frac{1}{\left|E(t)\right|}\left|E(t) \cap E(t-1)\right|,
\label{eq:sstep}
\end{equation}
where $E(t)$ is the set of edges of the network at time $t$, `$\cap$' is the intersection operator and the operator `$\left|\dots\right|$' gives the number of elements in the set.

\begin{figure}
\resizebox{1\columnwidth}{!}{%
\includegraphics{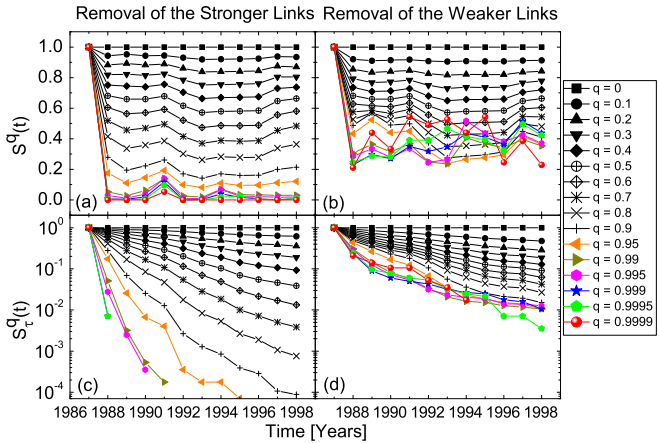}
}
\caption{Annual dynamics of the network similarity after we removed a fraction $q$ of the initial links. (a)~Single-step similarity probability, $S^{q}(t)$, of the networks after we remove the stronger links. (b)~Single-step similarity probability, $S^{q}(t)$, of the networks after we remove the weaker links. (c)~Multi-step similarity probability, $S_{\tau}^{q}(t)$, of the networks after we remove the stronger links. (d)~Multi-step similarity probability, $S_{\tau}^{q}(t)$, of the networks after we remove the weaker links.}
\label{fig:5}
\end{figure}

Accordingly, the \textit{multi-step similarity probability} at time $t$, after we remove a fraction $q$ of the initial network, is defined as:
\begin{eqnarray}
	S_{\tau}^{q}(t)&=&\frac{1}{\left|E(t)\right|}\left|E(t) \cap E(t-1)\dots \right. \nonumber\\ 
	&&\left. \dots \cap E(t-\tau+1) \cap E(t-\tau)\right|,
\label{eq:mstep}
\end{eqnarray}
where only the edges that are continuously present on the network after $\tau$ time steps are counted. Plots of the above quantities are shown at Figure~\ref{fig:5}. From these plots we can see that when we remove the weaker correlated links we are left with a small, stable, and very strongly connected network (Figs~\ref{fig:5}(b) and (d)), while when we remove the stronger correlated links we are left with a larger network that is not stable over time (Figs~\ref{fig:5}(a) and (c)).

Our results clearly suggest the presence of a nucleus of few strongly connected stocks in the stock market that form a stable structure over time. On the other hand, we see that the stocks that are not so strongly connected are those which form a much larger network (a representative network of the whole market). The presence of these noisy connections in the network makes it almost impossible to predict the price movement of one stock only by using information about the price movement of another.

\begin{table*}
\centering
\caption{Tick symbols of the stocks belonging to the communities shown in Fig~\ref{fig:2}. The symbols $\dag, \dag\dag, \ddag$ and $\ddag\ddag$ mark stocks that appear in more than one community.}
\label{tab:communities}
\begin{tabular}{lllllllll}
\hline\noalign{\smallskip}
 & (a) & (b) & (c) & (d) & (e) & (f) & (g) & (h) \\
\noalign{\smallskip}\hline\noalign{\smallskip}
1 & ALK & HIT & $\rm{HAL}^{\dag}$ & AA  & ABX & AHC &AIT &AIG \\
2 & AMR & HMC & HP  & AL  & AEM & ARC &BEL &AXP \\
3 & DAL & KYO & PKD & AR  & ASA & CHV &BGE &BMY \\
4 & U   & MC  & SLB & N   & BMG & $\rm{HAL}^{\dag}$ &$\rm{GE}^{\dag\dag}$  &CL  \\
5 & LUV & PIO & TDW & PD  & CDE & KMG &GTE &DIS \\
6 & FDX & SNE &     & RLM & HL  & MOB &HSY &EMR \\
7 & $\rm{GE}^{\dag\dag}$  & TDK &     &     & HM  & P   &JPM &FNM \\
8 &     &     &     &     & NEM & SC  &K   &$\rm{GE}^{\dag\dag}$  \\
9 &     &     &     &     &     & SLB &$\rm{KO}^{\ddag}$  &JNJ \\
10&     &     &     &     &     & TX  &MMM &$\rm{KO}^{\ddag}$  \\
11&     &     &     &     &     & UCL &$\rm{PG}^{\dag}$  &MER \\
12&     &     &     &     &     & $\rm{XON}^{\ddag\ddag}$ &SLE &MRK \\
13&     &     &     &     &     &     &T   &$\rm{PG}^{\dag}$  \\
14&     &     &     &     &     &     &UN  &SGP \\
15&     &     &     &     &     &     &USW &TY  \\
16&     &     &     &     &     &     &$\rm{XON}^{\ddag\ddag}$ &WMT \\
\noalign{\smallskip}\hline
\end{tabular}
\end{table*}
 
\section{Discussion}
\label{sec:3}

We found that correlation based networks show great tolerance to the removal of the stronger links since the network remains connected until we remove almost $99.9\%$ of its original links. If the removal is targeted to the weaker links, it results to a faster removal of nodes, but leaving the remaining network highly connected and highly clustered. This shows that the tolerance of the network to random and to intentional attacks on strong links comes mostly due to the connectivity provided by its weak links. This behavior is contrary to what happens in scale free networks, which show a large tolerance to random failures or attacks, but are highly vulnerable to intentional attacks~\cite{bb:Cohen,bb:AlbertJeongBarabasi2000,bb:Cohen2001,bb:Callaway2000}. 

Removal of the weak links from a correlation based network results in a somewhat shrinking of the network, but the properties of the remaining part are similar to the properties of the original network. This finding could explain why a financial market is not affected strongly by the small capitalization stocks that are usually weakly connected. On the contrary, it is strongly affected by the high capitalization stocks that are strongly connected most of the time. This behavior could not be explained by a scale free topology, because in that case a targeted attack to the strongly correlated nodes would result in a breakdown of the connections of the network.

Summarising, our results suggest that there is strong correlation between the topology of the network and the weights  of the links, which in our case is the correlation strength between the different stocks. The network disintegrates without a  sharp percolation transition when we sequentially remove either its weaker or its stronger correlated links. Since the network lacks a natural cutoff there is always an arbitrariness in the method one uses to filter out links of the original network, which might result in losing valuable information. However, since close to removing $99.5\%$ of the weak links we identify clear structures, we used this threshold as our parameter which indeed show the network's meaningful structure. 

We also find similar results by performing the same analysis using closing prices of a different market, the Athens Stock Exchange (ASE) for the period 1987 to 2004 (These results will be published elsewhere). This signifies that our findings are general and do not depend on the particular investigated system. However, we must keep in mind that correlation based networks are different from ordinary networks due to the fact that the link associated with each interaction is estimated starting from the statistical evaluation of the correlation coefficient. Therefore, one could use an estimation of reliability, such as the bootstrap technique~\cite{bb:TumminelloChaos}, to obtain reliability values for all links of the communities detected with the k-clique method, but such an analysis is beyond the scope of this paper.

\section{Acknowledgments}
\label{sec:4}

This work was partially supported by a European research NEST/PATHFINDER
project, DYSONET 012911, by the Greek General Secretariat for Research and Technology of the
Ministry of Development, PENED project 03ED840 and by the Israel Science Foundation.

\end{document}